# Anisotropic Phonon Transport, Thermal Expansion and Thermomechanics in Monolayer Td-WTe$_2$

Hongyi Sun[1], Qingfang Li[2*]

[1]National Laboratory of Solid State Microstructures, College of Physics, Nanjing University, Nanjing, 210093, China

[2]Department of Physics, Nanjing University of Information Science & Technology, Nanjing 210044, China

* Correspondence and requests for materials should be addressed to Qingfang Li (e-mail: qingfangli@nuist.edu.cn)

## Abstract:

Td-WTe$_2$, a typical member of two-dimensional layered transition metal dichalcogenides (TMDC), has recently drawn huge attentions due to its unique physical properties. In this paper, we elaborately investigate the phonon transport, thermal expansion and thermomechanics of monolayer Td-WTe$_2$ by using first-principles. We find the thermal properties of monolayer Td-WTe$_2$ are unique. The intrinsic lattice thermal conductivity is strongly anisotropic (a factor of 2) and very low (10 Wm$^{-1}$K$^{-1}$). But for the case of thermal expansion, the distinct anisotropy is not observed until at high temperature range. We find the anisotropic in-plane Young's moduli are stiffened by phonon excitations, while the Poisson's ratios along both directions are softened when temperature rises, showing opposite temperature effects on in-plane stiffness and in-plane strain-strain correlation. The mechanisms of these unique thermal properties are also discussed in this paper.

# 1. Introduction

In the last few decades, two-dimensional (2D) layered transition metal dichalcogenides (TMDC) have stimulated an intense research owing to their unique physical properties and potential applications.[1-5] Based on TMDCs, various of practical applications, such as field effect transistors (FETs),[6-8] thermoelectric devices[9] and energy storage,[10] have been explored. These applications make it imperative to investigate the thermal properties of these systems. For instance, the lattice thermal conductivity ($\kappa_l$) is critical in thermoelectric performance,[9,11] and the linear thermal expansion coefficients (LTEC) must be considered in heterostructure designs.[12-16]

Td-WTe$_2$, a typical member of TMDC, has recently drawn an immense attention due to its unique physical properties, such as unsaturated giant magnetoresistance,[17,18] superconductivity,[19-20] and the novel type-II weyl semimetal.[21] Inspired by these observations, many works begin to explore the properties of monolayer Td-WTe$_2$. It is reported that monolayer Td-WTe$_2$ preserves the semimetallic property with equal hole and electron carrier concentrations, which indicates that monolayer Td-WTe$_2$ may have the same extraordinary magnetoresistance effect as bulk, showing promising applications in nanostructured magnetic devices.[22] Qian *et al*. proposed that monolayer Td-WTe$_2$ is the most promising 2D TMDC for realizing topological FETs which can be rapidly switched off by electric field through a topological phase transition instead of carrier depletion.[23] E. Torun *et al*. reported the anisotropic and strain dependent electronic and optical properties of monolayer Td-WTe$_2$, which shows potential applications in nanoscale devices.[24] However, the study on the thermal properties is still in its infancy, despite they are critical in practical applications. For instance, monolayer Td-WTe$_2$ is expected as a potential thermoelectric material,[25] and its thermoelectric performance can be further improved by doping. According to classical theory, Td-WTe$_2$ should possess a low $\kappa_l$ due to its heavy atom mass and low Debye temperature.[26] But a detail knowledge of $\kappa_l$ is still lacking.

In this work, the thermal properties of monolayer Td-WTe$_2$, including phonon transport, phonon lifetimes, LTECs, temperature dependent Young's moduli and the Poisson's ratios are elaborately studied by using first-principles. Our simulations indicate that the thermal properties of monolayer Td-WTe$_2$ are unique. (i) The $\kappa_l$ is strongly anisotropic and very low (10 Wm$^{-1}$K$^{-1}$), and nanostructuring could be an effective way to further reduce the $\kappa_l$. (ii) But, unlike the strong anisotropy of $\kappa_l$ which is observed within the whole considered temperature range, the distinct anisotropy of LTEC is not observed until at high temperature range. (iii) Interestingly, the phonon excitations may stiffen the anisotropic in-plane Young's moduli, while the Poisson's ratios along both directions are softened as temperature increasing, showing opposite temperature effects on the in-plane stiffness and in-plane strain-strain correlation.

## 2. Method and Computational Details

All the first-principles calculations are performed based on density functional theory (DFT) as implemented in the Vienna ab-initio simulation package (VASP).[27-29] A cutoff energy of 400 eV is chosen with the exchange-correlation functional of generalized gradient approximation (GGA) of Perdew-Burke-Ernzerhof (PBE).[30] We use a uniform Γ-centered k-mesh of 18×10×1 to sample the Brillouin zone of monolayer Td-WTe$_2$ in the structure optimization and total energy calculation. The lattice constants and internal coordinates are optimized until the atomic forces become less than $10^{-4}$ eV/Å. To simulate the 2D property of monolayer Td-WTe$_2$, a vacuum space of at least 23 Å is included along the Z direction to minimize the interaction between the periodic images. Phonon dispersion curves are computed from the finite displacement method as implemented in the PHONOPY package.[31] A 4×4×1 supercell with a Γ-centered 5×3×1 k-mesh are chosen to calculate the harmonic interatomic force, and a 3×3×1 supercell with a Γ-centered 6×3×1 k-mesh for the calculations of anharmonic interatomic force which includes interactions up to the fourth nearest neighbors.

The intrinsic $\kappa_l$ is obtained by iteratively solving the linearized Boltzmann-Peierls transport equation as implemented in the ShengBTE code.[32] To ensure the conductivity convergence, we chose a 45×25×1 q-mesh. This program predicts material's $\kappa_l$ without any other adjustable parameters except for the basic information of the chemical structure.

To determine the thermal expansion and temperature dependent elastic moduli, Helmholtz free energy are calculated by using PHONOPY package[31] and quasiharmonic approximation (QHA):[13,14]

$$F(a_i, b_i, T) = E(a_i, b_i) + \sum_{\mathbf{q}j} \frac{\hbar \omega_{\mathbf{q},j}^i}{2} + k_B T \sum_{\mathbf{q}j} \ln\left(1 - \exp\left[-\frac{\hbar \omega_{\mathbf{q},j}^i}{k_B T}\right]\right), \quad (1)$$

where $a$ and $b$ are the lattice parameter, $i$ is the structure index, $E(a_i, b_i)$ is the ground state total energy of structure $i$, $\omega^i_{\mathbf{q},j}$ is the phonon frequency with the phonon mode index $j$ and phonon wave vector $\mathbf{q}$. The second and third terms are the contributions from zero point and vibrational energies. Here, the electronic contribution to free energy is neglected, as the electronic density of states of Td-WTe$_2$ at the Fermi level is quite small.[20] To obtain the lattice parameter dependent free energy function, thirty-six sets of ($a_i$, $b_i$) around the equilibrium lattice are considered between 0.98 and 1.03 $a_0$ ($b_0$). At each specified structure, we calculated the Helmholtz free energy with a temperature step of 1 K from 0 to 600 K. And a 324×180×1 q-point mesh is used. Then, at each temperature point, we interpolated the sets of free energy with spline function and obtained the equilibrium lattice constants at this temperature by minimizing the free energy.[33] Then, the thermal expansion coefficient (LTEC) and the temperature dependent elastic moduli are determined as follow:

$$\alpha(T) = \frac{1}{a_0(T)} \frac{da_0(T)}{dT}, \tag{2}$$

$$C_{\alpha\beta} = \frac{1}{Sh} \left. \frac{\partial^2 F}{\partial \varepsilon_\alpha \partial \varepsilon_\beta} \right|_T, \tag{3}$$

where $a_0(T)$ is the equilibrium lattice constant at temperature $T$, $\varepsilon_\alpha$ is the strain tensor with the standard Voigt notation, $S$ is the surface area at temperature $T$, $h$ is thickness of monolayer Td-WTe$_2$ (0.7 nm).[34] The Young's moduli and Poisson's ratios of monolayer Td-WTe$_2$ can be determined by:[35,36]

$$\begin{cases} E_x = \dfrac{C_{11}C_{22} - C_{12}^2}{C_{22}}, \quad E_y = \dfrac{C_{11}C_{22} - C_{12}^2}{C_{11}}, \\ \nu_{xy} = \dfrac{C_{12}}{C_{22}}, \quad \nu_{yx} = \dfrac{C_{12}}{C_{11}} \end{cases} \tag{4}$$

# 3. Results and Discussions

## 3.1 Lattice structure and vibrational properties

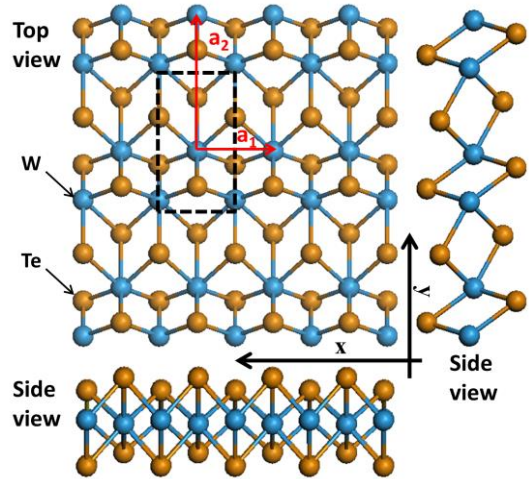

FIG. 1. The crystal structure of monolayer Td-WTe$_2$ in top and side views, and the Wigner-Seitz cell is shown by dashed lines.

The crystal structure of monolayer Td-WTe$_2$ viewed along three directions is displayed in Fig.1. Unlike most TMDCs which possess trigonal prismatic or monoclinic structures,[1] Td-WTe$_2$ adopts a distorted octahedral coordination (Td) orthorhombic structure. The octahedron of Te atoms in monolayer Td-WTe$_2$ is slightly distorted and the W atoms are displaced from their ideal octahedral sites, forming zigzag W-W chains along the *x* direction.[34] The distinct structural difference between *x* and *y* directions implies anisotropic in-plane properties. Our optimized lattice constants are a$_1$ = 3.5 Å, a$_2$ = 6.33 Å, which are in good agreement with previous studies.[22,24]

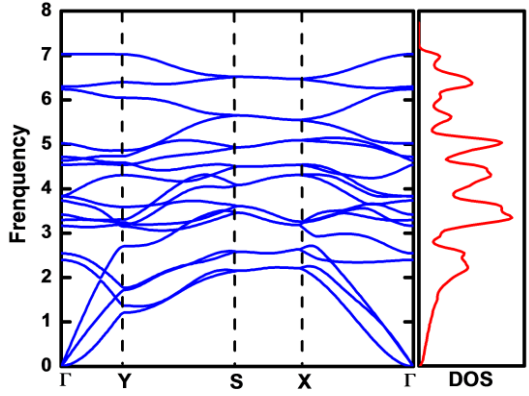

FIG. 2. The phonon dispersions and DOS of monolayer Td-WTe$_2$.

The phonon spectrum along high-symmetry lines together with the phonon density of states (DOS) are shown in Fig. 2. All the phonon modes of monolayer Td-WTe$_2$ are positive, indicating a dynamically stable configuration. Near the Γ point, the lowest acoustic branch, ZA-branch, is quadratic while the other two acoustic branches show linear, which is similar to that found in graphene[37] and other 2D TMDCs.[4] Due to the heavy atom mass and weak bonding stiffness, the total phonon dispersions only extend up to 7 THz which is about half of that in MoS$_2$.[12] Furthermore, unlike most H-phase TMDCs,[4] the Td-phase WTe$_2$ show gapless between the acoustic and optical phonon branches (as seen in DOS). This implies much more frequent interactions between acoustic modes and optical modes, resulting in a reduction in the lattice thermal transport.

### 3.2 Phonon transport properties

Based on the phonon spectrum and third order interatomic force, the $κ_l$ is determined by iteratively solving the linearized Boltzmann-Peierls transport equation. The results are shown in Fig. 3. As is shown in Fig. 3(a), the $κ_l$ of monolayer Td-WTe$_2$ is remarkably anisotropic and very low. The $κ_l$ along *x* direction (direction of W-W chains) is around half of that along *y* direction within the considered temperature range. At 300 K, the $\kappa_l$ along *x* direction is only 10 Wm$^{-1}$K$^{-1}$ which may be, to our knowledge, one of the lowest reported values of 2D materials. We note that the intrinsic $\kappa_l$ for 2H-MX$_2$ (M=Mo, W; X=S, Se) are all above 50 Wm$^{-1}$K$^{-1}$, those for 1T-MX$_2$ (M=Zr, Hf; X= S, Se) are all above 10 Wm$^{-1}$K$^{-1}$,[38] and those for h-BN and graphene are respectively more than 600 and 5000 Wm$^{-1}$K$^{-1}$.[39,40] The $\kappa_l$ along both directions decrease monotonously with the rise of temperature, which is attributed to the enhanced phonon-phonon scattering at higher temperature. Furthermore, we find the single-mode-relaxation time approximation (SMRTA) under-predicts the $\kappa_l$ along *x* (*y*) directions only by a factor of 1.24 (1.03) at 300 K, which is very different with that of graphene where the SMRTA under-predicts the $\kappa_l$ by more than a factor of 5 at 300 K.[41-43]

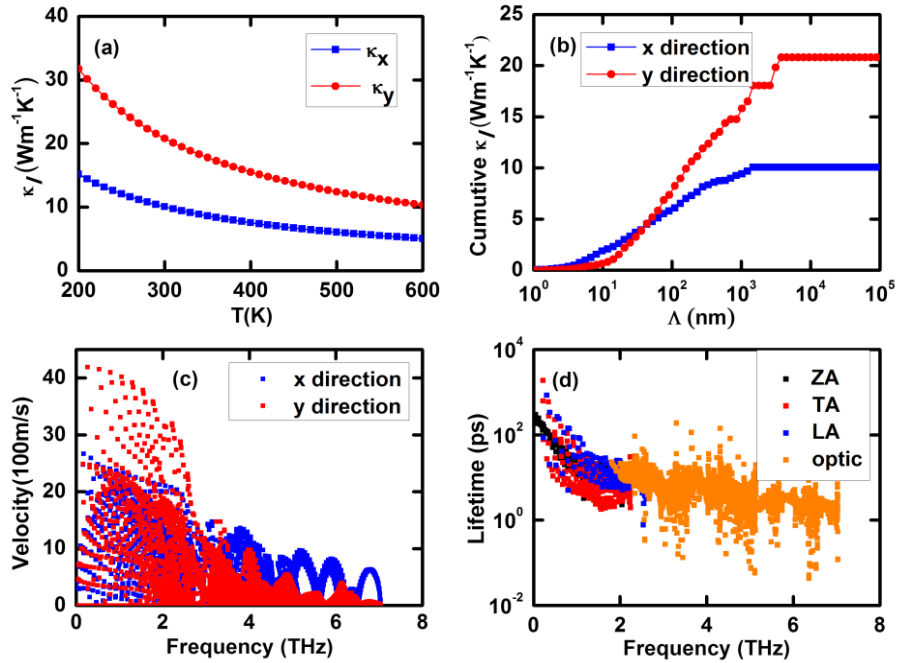

FIG. 3. (a) The temperature dependent lattice thermal conductivity of monolayer Td-WTe$_2$. (b) The cumulative thermal conductivity at 300 K. The predictions are connected by lines to guide the eye. (c) The group velocities. (d) The lifetimes of phonons at 300 K.

Moreover, to investigate the contributions from phonons with different mean free paths (MFP), we calculate the cumulative $\kappa_l$ along both $x$ and $y$ direction at 300 K. The cumulative $\kappa_l$ is determined by summing the thermal conductivity contributions of modes with MFPs up to $\Lambda$. The result is depicted in Fig. 3(b). As the $\Lambda$ increasing, the cumulative $\kappa_l$ keeps increasing at first, and then approaches plateau after $\Lambda$ reaches 1.5 μm and 3.7 μm for the $x$ and $y$ directions, respectively. Therefore, the heat is mainly carried by phonons with MFPs below 1.5 μm and 3.7 μm for the $x$ and $y$ directions, respectively. And as viewed from applications, the $\kappa_l$ can be significantly reduced when the characteristic lengths are smaller than 1.5 to 3.7 μm along the $x$ and $y$ directions.

To get insights into the properties of $\kappa_l$ given above, we further calculate the group velocities and phonon lifetimes of monolayer Td-WTe$_2$. The results are shown in Fig. 3(c) and (d). As is seen from Fig. 3(c), the velocities of acoustic modes along $y$ direction are significantly higher than that along $x$ direction, which is response for the anisotropic $\kappa_l$. The velocity of long-wave-limit acoustic branches along $x$ ($y$) direction is about 2.6 Km/s (4.2 Km/s) which is much small than those of MoS$_2$ (6.5 Km/s) and graphene (21.9 Km/s), implying a much lower $\kappa_l$. While for optical modes, the velocities along $x$ direction possess larger values than that along $y$ direction. Due to these special distributions of group velocities, the $\kappa_l$ contributions of different modes along $x$ direction can be very different from that along $y$ direction. For instance, the $\kappa_l$ contributions of ZA, TA, LA and optical modes along $x$ direction are 43.7%, 14.4%, 24.1% and 17.8% at 300 K respectively, and they are 13.7%, 20.3%, 54.1% and 11.9% along *y* direction. We note that the $\kappa_l$ contributions of different acoustic branches in monolayer Td-WTe$_2$ are very similar to those of phosphorenes and MoS$_2$.[44,45] But they are very different with that of graphene, where the contribution from ZA modes is above 80%.[41,46] In addition, the observed low $\kappa_l$ is also attributed to the short phonon lifetimes, as shown in Fig. 3(d). The lifetimes of ZA, TA and LA modes are mainly below 100 ps which are about four orders-of-magnitude smaller than that of graphene's ZA modes.[41,46] The short lifetimes of acoustic modes in monolayer Td-WTe$_2$ may mainly result from: (i) The narrow extension of acoustic dispersions which reduces the population of phonons with large velocities and simultaneously increases the number of three-phonon scattering processes, as more phonon modes are active at a given temperature. (ii) The gapless nature between the acoustic and optical branches which results in frequent scattering between acoustic modes and optical modes. (iii) The non-plane lattice structure which breaks the out-of-plane symmetry and further reduces the lifetime of ZA phonons.[41,46,47]

**3.3 Thermal expansion and temperature dependent elastic moduli**

We further investigated the thermomechanics of monolayer Td-WTe$_2$. Similar to the $\kappa_l$, the LTECs and temperature dependent moduli are related to the anharmonic interaction. But, unlike the high anisotropy of $\kappa_l$ which possesses a factor of up to 2 within the whole considered temperature range, the thermomechanics show anisotropic properties very differently. The details are below.

The calculated LTECs, grüneisen parameters, temperature dependent Young's moduli and Poisson's ratios are displayed in Fig. 4. As is seen from Fig. 4 (a), The LTECs of both directions are positive, except for very small negative thermal expansion around 0 K (max -0.46×$10^{-6}$ $K^{-1}$). The difference between LTECs along *x* and *y* directions is small until 150 K above which the difference increases with temperature, resulting in remarkable anisotropic property. The sharp change at low temperature is due to the small extension of phonon frequencies which allow phonons to be active at low temperature. We note that the LTECs of monolayer Td-WTe$_2$ are similar to that of other TMDCs,[13] but very different with graphene and h-BN where remarkable negative LTECs are observed.[14] The message given above can be useful in heterostructure designs based on monolayer Td-WTe$_2$.

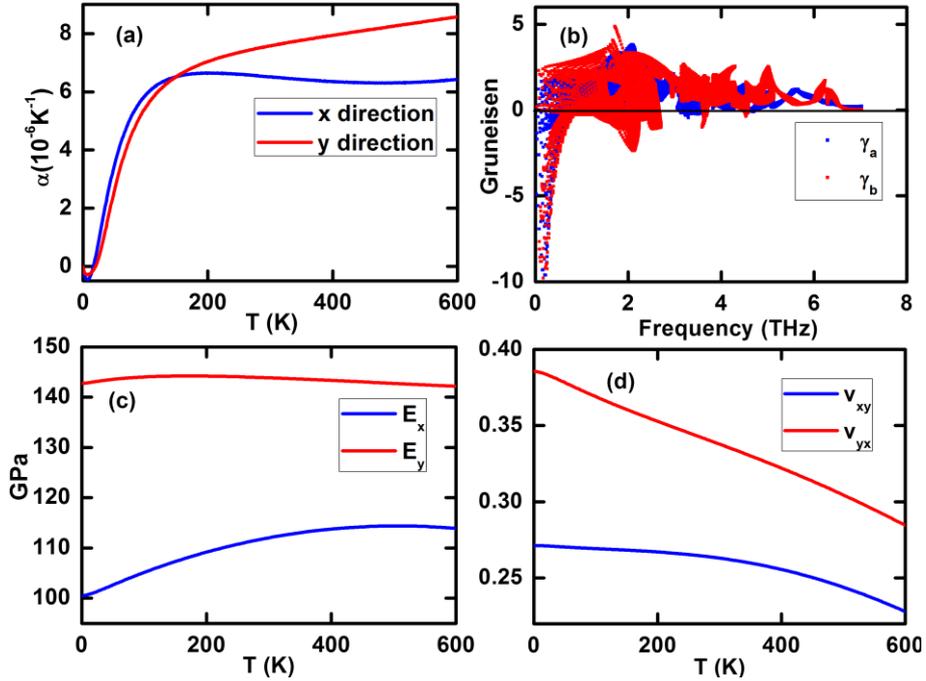

FIG. 4. (a) The temperature dependent LTECs of monolayer Td-WTe$_2$. (b) The grüneisen parameters $\gamma_a$ and $\gamma_b$. (c) The temperature dependent Young's moduli. (d) The temperature dependent Poisson's ratios.

The properties of LTECs can be understood by the grüneisen theory, where the LTECs of monolayer Td-WTe$_2$ can be expressed as:[48]

$$\begin{cases} \alpha_a = \dfrac{1}{V}\sum_{\mathbf{q}j} C_{\mathbf{q}j}[S_{11}\gamma_a(\mathbf{q},j) + S_{12}\gamma_b(\mathbf{q},j)] \\ \alpha_b = \dfrac{1}{V}\sum_{\mathbf{q}j} C_{\mathbf{q}j}[S_{12}\gamma_a(\mathbf{q},j) + S_{22}\gamma_b(\mathbf{q},j)] \end{cases}, \quad (5)$$

where $V$ is the volume, $C_{\mathbf{q}j}$ is the specific heat. $S_{11}$, $S_{12}$ and $S_{22}$ are the components of the elastic compliance tensor.[35] $\gamma_a(\mathbf{q},j)$ and $\gamma_b(\mathbf{q},j)$ are the generalized grüneisen parameter[48,49], and their values are displayed in Fig. 4(b). For a simply and qualitative analysis, we take $S_{11}$, $S_{12}$ and $S_{22}$ as constants with values at 0 K. According to our calculation, $S_{12}$ is negative, and $S_{11}$ and $S_{22}$ are about 3.68 and 2.59 times as large as the absolute value of $S_{12}$. For clarity, we choose the value of $S_{12}$ as the negative unit and express $S_{11}$ and $S_{22}$ in this unit. From Eq. (5), we get

$$\begin{cases} \alpha_a = \dfrac{1}{V}\sum_{\mathbf{q}j} C_{\mathbf{q}j}[3.68\gamma_a(\mathbf{q},j) + (-1)\gamma_b(\mathbf{q},j)] \\ \alpha_b = \dfrac{1}{V}\sum_{\mathbf{q}j} C_{\mathbf{q}j}[(-1)\gamma_a(\mathbf{q},j) + 2.59\gamma_b(\mathbf{q},j)] \end{cases}, \quad (6)$$

$$\alpha_b - \alpha_a = \dfrac{1}{V}\sum_{\mathbf{q}j} C_{\mathbf{q}j}[3.59\gamma_b(\mathbf{q},j) - 4.68\gamma_a(\mathbf{q},j)]. \quad (7)$$

As is seen from Eq. (6), the properties of $\alpha_a$ and $\alpha_b$ are largely determined by $\gamma_a(\mathbf{q},j)$ and $\gamma_b(\mathbf{q},j)$ respectively. Very near 0 K, the excited modes are mainly the low-frequency acoustic modes with negative $\gamma$ (Fig. 4(b)), therefore $\alpha_a$ and $\alpha_b$ possess negative values. As is seen from Eq. (7), and note that $\gamma_b(\mathbf{q},j)$ is overall comparable to $\gamma_a(\mathbf{q},j)$ at acoustic range, therefore the difference between $\alpha_a$ and $\alpha_b$ at low temperature range is small. At high temperature range, the increasing difference between $\alpha_a$ and $\alpha_b$ can be attributed to the excited high-frequency optical modes whose $\gamma_b(\mathbf{q},j)$ possess larger values than $\gamma_a(\mathbf{q},j)$, as shown in Fig. 4(b). In addition, we note that the large negative values in $\gamma_a$ and $\gamma_b$ are mainly from ZA modes, which is also found in other 2D materials, resulting from the membrane effect.[49]

Based on the calculated free energy, we obtained the temperature dependent Young's moduli and Poisson's ratios which respectively character the in-plane stiffness and strain-strain correlation between different directions. The results are shown in Fig. 4(c) and (d). As is seen from Fig. 4(c), the in-plane stiffness of monolayer Td-WTe$_2$ is remarkably anisotropic with considered temperature range. Interestingly, $E_x$ is significantly stiffened as $T$ increasing while $E_y$ keeps almost unchanged. This is different with that in graphene and MoS$_2$ where the in-plane stiffness are softened with $T$ increasing.[12,50] Since the thermal expansion generally decreases the curvature of the potential-energy surface and then softens the stiffness, we can conclude that the phononic excitations in monolayer WTe$_2$ have stiffened effects on $E_x$ and $E_y$. Because of the relation $\frac{v_{yx}}{v_{xy}} = \frac{E_y}{E_x}$, the Poisson's ratios of monolayer Td-WTe$_2$ along $x$ and $y$ directions are also very different (Fig. 4(d)). But unlike the temperature stiffened effect observed in $E_x$ and $E_y$, both the two Poisson's ratios decrease with $T$, indicating a weakened strain-strain correlation between $x$ and $y$ directions as temperature rises.

## 4. Conclusion

Using first-principles, we present an elaborate investigation of the phonon transport, thermal expansion, temperature dependent Young's moduli and Poisson's ratios of monolayer Td-WTe$_2$. We find the lattice thermal conductivity is strongly anisotropic and very low, which is attributed to the high-anisotropic-and-small-values group velocities and the short phonon lifetimes. Moreover, nanostructuring could be an effective way to further reduce the lattice thermal conductivity when its characteristic length is reduced to less than 1.5 to 3.7 μm. Unlike the anisotropy of lattice thermal conductivity, the distinct anisotropy of LTEC is not observed until at high temperature range. This is due to the special properties of the grüneisen parameters and elastic moduli of monolayer Td-WTe$_2$. Furthermore, we find the phonon excitations may stiffen the anisotropic in-plane Young's moduli, while the Poisson's ratios along both directions are softened when temperature rises, indicating opposite temperature effects on the in-plane stiffness and in-plane strain-strain correlation. Our studies provide insights into the thermal properties of monolayer Td-WTe$_2$ and should be useful in its future potential device applications at finite temperature.

## Acknowledgements

This work was supported by Natural Science Foundation of China (Grant Nos 11547030, 11525417, 11374142).